\documentclass[13pt]{article} 

\usepackage{color}
\usepackage{xcolor}
\usepackage{pdfsync}
\usepackage{amssymb}
\usepackage{amsmath,amsthm,amsfonts}  
\usepackage{graphicx}
\usepackage[breaklinks,colorlinks,backref]{hyperref}
\UseRawInputEncoding 
\DeclareMathOperator{\sgn}{sgn}

\begin{document}

\title{
{\Large \bf Multi-parametric  Nonlinear Generalization of Klein-Gordon: Real and Complex Fields}
}

\author{
M. A. Rego-Monteiro and E. M. F. Curado  \\
Centro Brasileiro de Pesquisas F\'\i sicas \\ 
Rua Xavier Sigaud 150, 22290-180 \\ Rio de Janeiro, RJ, Brazil \\
e-mail:   regomont@cbpf.br }  
\maketitle

\begin{abstract}
 
\indent

We construct a nonlinear multiparametric Klein-Gordon for complex and real fields with mass dimension depending on a real parameter $\alpha$ as $\delta = 2/(1+\alpha)$ where $\delta$ is the mass dimension of the fields. We show that there are three classes of generalized models, one class for complex fields and two different classes for real fields. All models in these three classes have travelling-wave solutions and satisfy the relativistic dispersion relation. Moreover, all models of the complex class and models of only one class of the two real classes recover the standard Klein-Gordon model. We also build the Lagrangian and the Hamiltonian for the three classes of models. The fields in the models of these three classes could in principle have the mass dimension varying from zero to one and this can allow us to construct interaction terms, other than $\lambda \Phi^4$, with coupling constants with positive or zero mass dimensions.
Furthermore, we also show that there is a subclass of equations in the complex class which has a Lorentzian soliton solution.

\end{abstract}

\begin{tabbing}

\=xxxxxxxxxxxxxxxxxx\= \kill

\\ Nonextensive thermostatistics; Renormalization; Solitons; Standard-Model Extension

 \\

\end{tabbing}

\newpage

\section{Introduction}
\label{intro}

In ref.\cite{q-3eq} it was presented a nonlinear generalization of the three main physics equations, notably Schr\"odinger, Klein-Gordon and Dirac equations in such a way that the standard linear equations are obtained in the limit $q \rightarrow 1$ and in all cases, the well-known relativistic dispersion relation is preserved. This parameter $q$ was introduced by imposing that these equations have as solution a generalization of the exponential function known as $q$-exponential function which appear in a variety of systems in the context of non-extensive statistical mechanics \cite{tsallisbook,tsallisreviewbjp09}.
Since the work in ref.\cite{q-3eq} was presented there has been a great deal of activity regarding its development and application \cite{jmp,pennini,rocca,toranzo,tsalliso,arplastino1,alves,aplastino1, arplastino2, aplastino2,bountis1, nobre1,arplastino3,nobre2, aplastino3, khosropour, ernesto, plastinosouza2014, lorentziansolitonqschroedinger, aplastinorocca4, meradaouachria, JZamoraMCRocca,Carvalho}. In particular,  interesting developments of the $q$-Klein-Gordon in ref.\cite{q-3eq} can be found in references \cite{arplastino2, aplastino2, arplastino3,aplastino3} and \cite{aplastinorocca4, meradaouachria, JZamoraMCRocca,Carvalho}.

Quantum field Theory is the basic framework used to describe particle physics \cite{greiner}. Its ideas were also used for condensed matter \cite{qftcm}, in particular in the theory of metals, superconductivity, the quantum Hall effect, among others. However, most interacting models in quantum field have infinities and just a few of these models have infinities that could be removed by redefining a finite number of physical quantities by a procedure known as renormalization. Thus, there is a limited number of quantum field models which are renormalized and it would be interesting if one could enlarge this class of renormalized models.

From perturbation theory, in the non-renormalized models the mass dimensions of the coupling constants are negative. 
An interaction-term is given by a combination of the coupling constant times a product of fields, such that its total mass dimension has to be equal to the value of the space-time. Thus, given the mass dimension of a field there will be a limited number of interaction-terms such that the mass dimension of the coupling constant is non-negative.  In the nonlinear generalized field models in ref.\cite{q-3eq} the mass dimensions of the fields are the same as one finds for their standard linear models, as a result the interaction-terms which could be renormalizable are the same in both cases. Then, in order of enlarging the class of possible interactions one could generalize the models in \cite{q-3eq} such that the mass dimension of the fields are different from one and analyze the consequences of introducing new interaction-terms with non-negative mass dimension coupling constants by such change in the mass dimension of the fields. The enlargement of the class of renormalized field theories, could be of interest, for instance, in the studies of relativity violations in the effective field theory called the Standard-Model Extension where some operators of interest are non-renormalized \cite{nonminimalsme}. 

In this paper, we present a nonlinear generalization of the $q$-Klein-Gordon in \cite{q-3eq} where the mass dimension of the field is $2/(1+\alpha)$ where $\alpha$ is a real constant. For $\alpha < 1$, as discussed above, we are allowed to construct models with interaction terms different from the standard ones, which are candidates to be renormalized models. By imposing that the generalized nonlinear equation has travelling-wave solutions which recovers the standard case and preserve the relativistic dispersion relation, we show that there are three classes of nonlinear generalized equations, one class for complex fields and two different classes for real fields. The  equations of the class for complex fields and the equations of one of the class for real fields recover in a limit the standard undeformed Klein-Gordon equation (see for instance \cite{greiner}). Instead the equations of the second class for real fields does not go at any limit to to the standard Klein-Gordon equation. It is also shown that there is a subclass of equations in the complex class which has a Lorentzian soliton solution. It is interesting to notice that Lorentzian solitary waves were already found in structures with photonic band gaps \cite{contitrillo1,contitrillo2} and also in internal water waves \cite{benjamin, ono}. Recently, it was shown in \cite{lorentziansolitonqschroedinger} a Lorentzian soliton in the $q$-Schr\"odinger equation in \cite{q-3eq}.

In section \ref{review} we present a brief review of the $q$-Klein-Gordon in \cite{q-3eq}, in section \ref{generalization} we introduce the generalized nonlinear Klein-Gordon equation. In subsection \ref{complex} we solve the generalized equation to the complex case. In subsection \ref{solutionreal} we solve to the real case and show that there are two different classes of solutions. 
In section \ref{lagangian/hamiltonian} we present the Lagrangian and Hamiltonian for the three classes of solutions given in subsections \ref{complexlagangian} - \ref{real2lagangian}. In section \ref{lorentzian}, we show that a subclass of the class of complex solutions to the nonlinear generalized Klein-Gordon equation has a spatial one-dimensional Lorentzian solitonic solution. Finally, in section \ref{final} we present our conclusions and the perspectives of this present work.

\section{Brief Review of a $q$-Klein-Gordon}
\label{review}
In this section we are going to present a brief review of the $q$-Klein-Gordon equation which appeared in \cite{q-3eq}. We start introducing the generalization of the standard exponential function called $q$-exponential given as 
\begin{equation}
\label{qexp}
e_q (x) \equiv \left[ 1+(1-q) x  \right] ^{1/(1-q)} , \,\, q \, \epsilon \, \Re ,
\end{equation}
where $e_{q} (x) \rightarrow \exp (x)$, as $q \rightarrow 1$, and $\exp(x)$ is the standard exponential function, is an important function in non-extensive statistical mechanics and is present in the description of several physical systems  \cite{tsallisbook, tsallisreviewbjp09}. 

The $q$-Klein-Gordon in \cite{q-3eq} may be written as
\begin{eqnarray}
\label{qkgprl}
\partial_\mu \partial^\mu   \Phi(x)+q c^{2(1-q)} m^{2 (2-q)}\, \Phi(x)^{2q-1}=0
\end{eqnarray}
where $c$ in the above equation is a constant and $\Phi(x) \equiv \Phi(\vec{x},t)$. The constant $c$ in eq.(\ref{qkgprl}) is connected to the constant $\Phi_0$ introduced in \cite{q-3eq} by the relation $c^{2(1-q)} m^{2\delta(1-q)} = \Phi_{0}^{2(1-q)}$. It is easy to see that $\Phi(z) = m^\delta c \, e_q(z)$, for $\delta = 1$ and $z=i(\omega t - \vec k. \vec x)$, is a solution to eq.(\ref{qkgprl}) where the relativistic dispersion relation $\omega^2 = k^2 + m^2$ is satisfied.

In \cite{jmp} an auxiliary field $\Phi_2(x)$ was introduced in order to construct the Lagrangian density of eq.(\ref{qkgprl}). This Lagrangian density can be written as
\begin{eqnarray}
\label{densityqkgprl}
\mathcal  L (\Phi_1(x),\Phi_2(x)) =     \partial_\mu \Phi_1(x) \partial^\mu \Phi_2(x)  
- \kappa \Phi_1(x)^{2 q - 1} \Phi_2(x) + C. C.
\end{eqnarray}
where $\kappa =q c^{2(1-q)} m^{2 + 2 (1-q)}$, $\Phi_1(x)$ is the field $\Phi(x)$ in eq.(\ref{qkgprl}) and $C.C.$ represents the complex conjugate terms of the previous terms. We also see that the mass dimension of the fields $\Phi_{1,2}(x)$ is $(\nu-2)/2$ where $\nu$ is the space-time dimension. The Euler-Lagrange equations for the field $\Phi_2(x)$ gives eq.(\ref{qkgprl}) and the Euler-Lagrange equations for the field $\Phi_1(x)$ gives
\begin{eqnarray}
\label{auxeq0}
 \partial_\mu \partial^\mu \Phi_2(x) + 
  + \kappa (2q-1)  \Phi_1(x)^{2q-2}  \Phi_2(x) =0
\end{eqnarray}
Differently from eq.(\ref{qkgprl}), this equation is linear for the auxiliary field $\Phi_2(x)$ and to solve it we must introduce the solution of $\Phi_1(x)$ from eq.(\ref{qkgprl}) which is $\Phi_1(z) = c \, m \, e_q(z)$. Thus, introducing this value of $\Phi_1(z)$ the solution to eq.(\ref{auxeq0}) is $\Phi_2(z) = k_1 e_q(z)^{1-2q}+k_2 e_q(z)^{-1+2q}$  \cite{jmp} .

The Hamiltonian for this case can be computed. The canonically conjugate momenta to tis case $\Pi_{1,2} = \partial \mathcal L/\partial \dot\Phi_{1,2}$ is given by $\Pi_1 = \dot\Phi_2 $ and $\Pi_2 =  \dot\Phi_1$. Thus the Hamiltonian density $\mathcal H = \Pi_1 \dot\Phi_1 + \Pi_2 \dot\Phi_2 - \mathcal L$ is
\begin{eqnarray}
\label{hamiltonian1}
\mathcal H = \Pi_1(x) \Pi_2(x)   + \vec{\nabla} \Phi_1(x) . \vec{\nabla} \Phi_2(x)  
+ \kappa \Phi_1(x)^{2q-1} \Phi_2(x) + C.C.
\end{eqnarray}

\section{The Generalized Nonlinear Klein-Gordon and Its Solution}
\label{generalization}

In this section we are going to generalize the $q$-Klein-Gordon model which appeared in \cite{q-3eq}. The field in this $q$-Klein-Gordon model  has mass dimension one and in what follows we are going to  generalize this equation such that the field has mass dimension depending on a real parameter $\alpha$.  As was already mentioned,  according to perturbation theory the mass dimension of a field in a field theory model give us the possible renormalizable interactions. 

Let us consider the following nonlinear multi-parametric generalization of the Klein-Gordon equation in an arbitrary space-time dimension:
\begin{equation}
\label{ansatz1}
\partial_\mu \partial^\mu   \Phi(x)^\alpha  +  a_1 \Phi(x)^{\alpha -2} \partial_\mu   \Phi(x) \partial^\mu    \Phi(x) + a_2  m^\beta  \Phi(x)^\gamma = 0  
\end{equation}
where $\alpha$, $\beta$, $\gamma$ and $a_{1,2}$ are arbitrary parameters. By a simple inspection we see that for $\alpha =1$, $\gamma =1$, $\beta =2$, $a_1 = 0$ and $a_2 = 1$ we recover the well-known standard Klein-Gordon equation.

Our strategy to study this nonlinear generalized equation in eq.(\ref{ansatz1}) is to search for relativistic traveling wave solutions to this equation respecting the relativistic dispersion relation. For this purpose, we define new variables $z_1  \equiv  k_\mu x^\mu = k^0 x^0 - \vec k . \vec x $ for the real case and $z_2  \equiv  i k_\mu x^\mu = i(k^0 x^0 - \vec k . \vec x)$ for the complex one. In terms of these new variables we obtain for eq.(\ref{ansatz1}):
\begin{equation}
\label{ansatz2}
\pm  k^2 \left [ \frac{d^2}{d z_{1,2}} \Phi(z_{1,2})^\alpha  + a_1 \Phi(z_{1,2})^{\alpha-2} \Phi^ \prime (z_{1,2})     \right ] + a_2 m^\beta \Phi(z_{1,2})^\gamma = 0
\end{equation}
where the plus sign comes together with variable $z_1$ and minus sign with variable $z_2$. Calling $\Phi(z_{1,2}) \equiv m^\delta c f(z_{1,2})$ where $\delta$ is the mass dimension of $\Phi(z)$, $c$ is a dimensionless constant and $f(z)$ is a dimensionless function of $z$. Thus, using this new definition we get:
\begin{equation}
\label{ansatz3}
k^2 \pm \frac{a_2 c^{\gamma - \alpha} f(z_{1,2})^{\gamma -\alpha +2}}{\alpha f(z_{1,2}) f^{\prime \prime}(z_{1,2}) + \left[   a_1 + \alpha (\alpha -1) \right]  f^{\prime}(z_{1,2})^2 } m^{\beta + \delta (\gamma - \alpha)} = 0
\end{equation}
where the plus and minus sign are for the real and complex cases respectively. Then, in order to have the relativistic dispersion relation for a free particle we must have
\begin{equation}
\label{EMR}
\frac{a_2 c^{\gamma - \alpha} f(z_{1,2})^{\gamma -\alpha +2}}{\alpha f(z_{1,2}) f^{\prime \prime}(z_{1,2}) + \left[   a_1 + \alpha (\alpha -1) \right]  f^{\prime}(z_{1,2})^2 } = \mp1
\end{equation}
with the positive sign for the complex case, the negative sign for the real one and $\beta + \delta (\gamma-\alpha)=2$. From eq.(\ref{ansatz1}) we can deduce also two additional relations between the coefficients. Taking into account that the action is dimensionless and supposing that the second field we add in the Lagrangian has the same mass dimension we have from the kinetic term  
$\delta= (\nu-2)/(1+\alpha)$
and from the mass term 
$\delta(1+\gamma)+\beta=\nu $
where $\nu$ is the space-time dimension. These last two equations give 
\begin{equation}
\label{beta}
\beta =2-\delta(\gamma-\alpha)
\end{equation}

The standard Klein-Gordon equation is obtained from $\alpha=\gamma=a_2=1$ and $a_1=0$ thus eq.(\ref{ansatz3}) for the real case is $k^2 + \frac{f^2}{f f^{\prime \prime}} m^2=0$ which has $\cos(z)$ or $\sin(z)$ as solution. For the complex case where the sign is negative we easily see that the solution is the exponential.

\subsection{Solution to the Complex Case}
\label{complex}

In order to obtain a complex traveling wave solution to eq.(\ref{ansatz1}), i.e, a function $f(z_2)$, and the conditions for the coefficients appearing in this equation satisfying the relativistic dispersion relation we must solve the eq.(\ref{EMR}) for the plus sign. One possible solution to this equation with plus sign is the well-known $q$-exponential given in eq.(\ref{qexp}). In this case, for $f(z) = e_q(z)$ we have $f(z) f^{\prime \prime} (z) =q e_q(z)^{2q}$, $f^{\prime}(z)^2 = e_q(z)^{2q}$ and $f(z)^{\gamma-\alpha+2} = e_q(x)^{\gamma-\alpha+2}$. If 
\begin{equation}
\label{relation1}
\gamma-\alpha+2=2q 
\end{equation}
eq.(\ref{EMR}) becomes independent of $z$, the $q$-exponential is a solution to it and we get the following relation among the coefficients
\begin{equation}
\label{relation2}
a_2 c^{2(q-1)} = \alpha q +a_1+\alpha(\alpha-1)
\end{equation}
In summary, we have found a complex solution to eq.(\ref{ansatz1}) as $\Phi(z) = m^\delta c \, e_q(z)$, where the coefficients are given by eqs.(\ref{beta}-\ref{relation2})  where the mass dimension of the field $\Phi(z)$ is  $\delta= (\nu-2)/(1+\alpha)$. The standard Klein-Gordon is obtained in the limit $a_1 \rightarrow 0;\, \alpha \rightarrow 1;\, a_2 \rightarrow 1;\, \beta \rightarrow 2$ and the $q$-Klein-Gordon which appeared in \cite{q-3eq} is obtained when $\alpha=\delta=1$ and $a_1=0$.

\subsection{Solution to the Real Case}
\label{solutionreal}

In the real case the equation to be solved is
\begin{equation}
\label{real}
a_2 c^{\gamma - \alpha} f(z)^{\gamma -\alpha +2}+\alpha f(z) f^{\prime \prime}(z) + \left[   a_1 + \alpha (\alpha -1) \right]  f^{\prime}(z)^2  = 0
\end{equation}
As in the previous case the solution is given by $\Phi(z)= c m^\delta f(z)$  and we choose for the real case we take the following ansatz $f(z) = \cos(bz)^\theta$. Substituting this ansatz in eq.(\ref{real}) we obtain the following equation:
\begin{eqnarray}
\label{real1}
a_2 c^{\gamma-\alpha} \cos(bz)^{\theta(\gamma-\alpha+2)} + \left[ -\alpha \theta b^2 + (a_1 + \alpha^2)\theta^2 b^2  \right] \cos(bz)^{2\theta-2} \nonumber  \\
 - (a_1+\alpha^2) \theta^2 b^2 \cos(bz)^{2\theta}=0 \,\, .
\end{eqnarray}
Notice that, the coefficient $a_2$ which appears in the first term of the above equation is present in the mass term of eq.(\ref{ansatz1}). Thus, in order to have a solution with non-zero mass this term should contribute for the second or third terms in eq.(\ref{real1}). Then, for non-zero mass solutions we have two different cases to consider, namely: case (I) $\theta (\gamma-\alpha +2) = 2\theta$ and case (II) $\theta (\gamma - \alpha + 2)= 2 \theta -2$. The same equations appears if we take the ansatz $f(z) = \sin(bz)^\theta$ instead of the previous one. 

\subsubsection{Real - Case (I)}
\label{real1.1}

In case 1, since $\theta \neq 0$ we have $\alpha = \gamma$ and we get the following equations:
\begin{eqnarray}
\label{eqreal1.1}
-\alpha \theta b^2 + (a_1+\alpha^2) \theta^2 b^2 =0   \\
\label{eqreal1.2}
-a_2 c^{\gamma -\alpha} + (a_1+\alpha^2) \theta^2 b^2=0
\end{eqnarray}
Since $\alpha = \gamma$ the constant $c$ is arbitrary and as $\beta =2-\delta(\gamma-\alpha)$ we have $\beta = 2$. Moreover, as $b \neq 0$ eq.(\ref{eqreal1.1})  implies that $a_1=\alpha(1-\alpha \theta)/\theta$ and from eq.(\ref{eqreal1.2}) we get $a_2 = \alpha \theta b^2$.

Notice that, the limit where $b \, \textnormal{,}\alpha \, \textnormal{and} \, \theta \rightarrow1$ reproduces the standard Klein-Gordon equation. Moreover, we antecipate that a further restriction to the possible values of $\theta$ will appear when we consider the Lagrangian density of this case.

\subsubsection{Real - Case (II)}
\label{real2.1}

The non-zero mass condition for case 2 give us $\alpha-\gamma = 2/\theta$ which leads to $\beta = 2+2\delta/\theta$. The equations in this case are $a_2 c^{\gamma-\alpha}-\alpha \theta b^2 + (a_1 + \alpha^2)\theta^2 b^2=0$ and $(a_1+\alpha^2) \theta^2 b^2=0$, which gives trivially
\begin{eqnarray}
\label{eqreal2.1}
a_1&=&-\alpha^2   \\
\label{eqreal2.2}
a_2&=&c^{\alpha-\gamma} \alpha \theta b^2
\end{eqnarray}

\section{Lagrangian and Hamiltonian to the Generalized Nonlinear Klein-Gordon Equation}
\label{lagangian/hamiltonian}

In this section we are going to derive the Lagrangian and Hamiltonian densities for the complex and real nonlinear generalized Klein-Gordon equations deduced in the previous section. 

\subsection{The Complex Case: Lagrangian and Hamiltonian}
\label{complexlagangian}

In the cases of the $q$-Schr\"odinger and complex $q$-Klein-Gordon equations presented in \cite{q-3eq} it was shown in \cite{qepl} and in \cite{jmp} that in order to derive a simple Lagrangian density to these complex equations it was convenient to introduce an additional auxiliary field. The same is true for the complex case under consideration. We then introduce two independent fields in the general case which correspond in the limit for the standard Klein-Gordon to the field and its complex conjugate. Let us consider the following Lagrangian density
\begin{eqnarray}
\label{complexlagrangian1}
\mathcal L (\Phi_1(x),\Phi_2(x)) &=&  \alpha  \Phi_1(x)^{\alpha-1} \partial_\mu \Phi_1(x) \partial^\mu \Phi_2(x) - a_1 \Phi_1(x)^{\alpha-2} \Phi_2(x)  \partial_\mu   \Phi_1(x) \partial^\mu   \Phi_1(x) \nonumber \\
&-& a_2 m^\beta \Phi_1(x)^\gamma \Phi_2(x) + C. C.
\end{eqnarray}
where $C.C.$ in the above equation means the complex conjugate of the previous terms in the equation. The values of $a_{1,2}$ and $\beta$ are given in this case by eqs.(\ref{beta}-\ref{relation2}).

Let us compute the Euler-Lagrange equation for the field $\Phi_2$. We have:
\begin{eqnarray}
\label{elphi2}
\frac{\partial \mathcal L}{\partial \Phi_2} &=& -a_1 \Phi_1^{\alpha-2} \partial_\mu\Phi_1 \partial^\mu \Phi_1 - a_2 m^\beta \Phi_1^\gamma  \\
\partial_\mu \frac{\partial \mathcal L}{\partial_\mu \Phi_2} &=& \alpha(\alpha-1) \Phi_1^{\alpha-2} \partial_\mu \Phi_1 \partial^\mu \Phi_1 + \alpha \Phi_1^{\alpha-1} \partial_\mu \partial^\mu \Phi_1
\end{eqnarray}
These two equations together give us the generalized nonlinear Klein-Gordon equation in eq.(\ref{ansatz1}) which we have solved in subsection (\ref{complex}). Moreover, the Euler-Lagrange for the complex field $\Phi_2^{\star}$ gives the complex conjugate of equation in eq.(\ref{ansatz1}). We must now compute the equation generated by the Euler-Lagrange for the field $\Phi_1$. From the above Lagrangian in eq.(\ref{complexlagrangian1}) we get
\begin{eqnarray}
\label{elphi1}
\frac{\partial \mathcal L}{\partial \Phi_1} = \alpha(\alpha-1) \Phi_1^{\alpha-2} \partial_\mu \Phi_1 \partial^\mu \Phi_2 - a_1 (\alpha-2) \Phi_1^{\alpha-3} \partial_\mu \Phi_1 \partial^\mu \Phi_1 \Phi_2      -a_2 \gamma m^\beta \Phi_1^{\gamma-1} \Phi_2   \\
\partial_\mu\frac{\partial\mathcal L}{\partial \partial_\mu \Phi_1} = \alpha \Phi_1^{\alpha-1} \partial_\mu \partial^\mu \Phi_2 - 2 a_1 \Phi_1^{\alpha-2} \partial_\mu 
\partial^\mu \Phi_1 \Phi_2 -    
\left[ 2a_1-\alpha(\alpha-1)  \right] \Phi_1^{\alpha-2} \partial_\mu \Phi_1 \partial^\mu \Phi_2  \nonumber \\
-2a_1(\alpha-2) \Phi_1^{\alpha-3} \partial_\mu\Phi_1 \partial^\mu\Phi_1 \Phi_2
\end{eqnarray}
Thus, the Euler-Lagrange equation for the field $\Phi_1$ is
\begin{eqnarray}
\label{auxeq}
\alpha \Phi_1^{\alpha-1} \partial_\mu \partial^\mu \Phi_2 - 2 a_1 \Phi_1^{\alpha-2} \partial_\mu \Phi_1 \partial^\mu \Phi_2 + \left[ -(\alpha-2) a_1 \phi_1^{\alpha-3} \partial_\mu \Phi_1 \partial^\mu \Phi_1 \right. \nonumber \\
\left. - 2a_1\Phi_1^{\alpha-2}  \partial_\mu \partial^\mu \Phi_1 + a_2 \gamma m^\beta \Phi_1^{\gamma-1} \right] \Phi_2 =0
\end{eqnarray}
Notice that, differently from the nonlinear eq.(\ref{ansatz1}), the above eq.(\ref{auxeq}) is a linear equation for the field $\Phi_2$ with coefficients dependent on the field $\Phi_1$. 

In order to solve eq.(\ref{auxeq}) it is convenient to rewrite it as:
\begin{eqnarray}
\label{auxeq1}
\alpha \partial_\mu \partial^\mu \Phi_2-2a_1 \Phi_1^{-1} \partial_\mu \Phi_1 \partial^\mu \Phi_2 + \left[ -(\alpha-2) a_1 \Phi_1^{-2} \partial_\mu\Phi_1 \partial^\mu \Phi_1  \right.  \nonumber \\ 
\left. -2a_1 \Phi_1^{-1} \partial_\mu \partial^\mu \Phi_1 + a_2  \gamma m^\beta   \Phi_1^{\gamma-\alpha}       \right] \Phi_2 =0
\end{eqnarray}
We should use the solution $\Phi_1(z) = m^\delta c \, e_q(z)$, eqs.(\ref{beta}-\ref{relation2}), where the mass dimension of the field $\Phi(z)$ is  $\delta= (\nu-2)/(1+\alpha)$. We see in eq.(\ref{auxeq1}) that the constants of $\Phi_1$ cancel in all terms, except in the last term of this equation where we have the contribution $a_2 c^{\gamma -\alpha} \gamma m^{\beta + \delta(\gamma-\alpha)}=a_2 c^{2q(q-1)} \gamma m^2$. Using that, $\Phi_1^{-1}(z) \partial_\mu \Phi_1(z)= i k_\mu e_q(z)^{q-1}$, $\Phi_1(z)^{-2} \partial_\mu \Phi_1(z) \partial^\mu \Phi_1(z)= -k^2 e_q(z)^{2q-2} $, $\Phi_1(z)^{-1} \partial_\mu \partial^\mu \Phi_1(z) = -q k^2 e_q(z)^{2q-2}$, $m^\beta \Phi_1(z)^{\gamma-\alpha} = c_1^{2q-2} m^2 e_q(z)^{2q-2}$ and $\Phi_2(z) = N_2 g(z)$ eq.(\ref{auxeq1}) can be written as
\begin{eqnarray}
\label{auxeq2}
-\alpha g^{\prime \prime}(z) + A e_q(z)^{q-1} g^\prime(z)+B e_q(z)^{2q-2} g(z)=0
\end{eqnarray}
where $A=2a_1$ and $B=  \left[ \alpha+2(q-1)\right]  \left[ 2 a_1 + \alpha (\alpha + q -1)     \right]$. 
We simply verify that the solution to eq.(\ref{auxeq2}) is $g(z)= \kappa_1 e_q(z)^{r_1} + \kappa_2 e_q(z)^{r_2}$ with $r_{1,2}$ and
\begin{equation}
\label{solutionphi2}
\Phi_2(z) = c_2 m^\delta \left [ \kappa_1 e_q(z)^{r_1} + \kappa_2 e_q(z)^{r_2} \right ]
\end{equation}
where 
\begin{eqnarray}
\label{rvalue1}
r_1 &=& 2 - 2 q - \alpha  \\
\label{rvalue2}
r_2 &=& \left(2 a_1 - \alpha + q \alpha + \alpha^2 \right)/\alpha.
\end{eqnarray}
The standard Klein-Gordon case is when $q \rightarrow 1$, $\alpha \rightarrow 1$ and $a_1=0$ giving $A=0$ and $B=1$. In this case, we obtain for $\kappa_2=0$ the standard solution $g(z) \rightarrow \exp(-z) = \exp(z)^\dagger$. Moreover, the case in \cite{jmp} is obtained for $\alpha \rightarrow 1$ and $a_1=0$ which leads us to 
$r_1=1-2q$  and
$r_2=q$.

It is important to observe that one could perform the transformation $\Psi(x) = \Phi(x)^\alpha$ in eq.(\ref{ansatz1}) obtaining a classically equivalent equation in terms of the field $\Psi(x)$. From the Lagrangian of this transformed equation we see that the mass dimension of $\Psi(x)$ is one. But in spite of these two equations are classically equivalent, their quantum systems can be completely different since we must consider the Jacobian of the transformation between  $\Phi(x)$ and $\Psi(x)$. This Jacobian is responsible for introducing the quantum differences between the two models (see for instance \cite{fujikawa}).

Let us compute the Hamiltonian for this case. The canonically conjugate momenta to this case $\Pi_{1,2} = \partial \mathcal L/\partial \dot\Phi_{1,2}$ is given by $\Pi_1 = \alpha \Phi_1^{\alpha-1} \dot\Phi_2 - 2 a_1 \Phi_1^{\alpha-2} \Phi_2 \dot\Phi_1$ and $\Pi_2 = \alpha \Phi_1^{\alpha-1} \dot\Phi_1$. Thus the Hamiltonian density $\mathcal H = \Pi_1 \dot\Phi_1 + \Pi_2 \dot\Phi_2 - \mathcal L$ is
\begin{eqnarray}
\label{hamiltonian1}
\mathcal H = \alpha \Phi_1^{\alpha-1} \dot\Phi_1 \dot\Phi_2  - a_1 \Phi_1^{\alpha-2} \Phi_2 (\dot\Phi_1)^2 + \alpha \Phi_1^{\alpha-1} \vec{\nabla} \Phi_1 . \vec{\nabla} \Phi_2  \nonumber \\
- a_1 \Phi_1^{\alpha-2} \Phi_2 (\vec{\nabla} \Phi_1)^2 + a_2 m^\beta \Phi_1^\gamma \Phi_2 + C.C.
\end{eqnarray}
It is easy to see that
\begin{eqnarray}
\label{relationsmomenta}
\frac{1}{\alpha} \Phi_1^{-\alpha+1} \Pi_1  \Pi_2 &=& \alpha \Phi_1^{\alpha-1} \dot\Phi_1 \dot\Phi_2 -2 a_1 \Phi_1^{\alpha-2} \Phi_2 (\dot\Phi_1)^2 \\
\frac{a_1}{\alpha^2} \Phi_1^{-\alpha} \Phi_2 \Pi_2^2 &=& a_1 \Phi_1^{\alpha-2} \Phi_2 (\dot\Phi_1)^2 
\end{eqnarray}
With these above equations we get:
\begin{eqnarray}
\label{hamiltonian1.1}
\mathcal H &=& \frac{1}{\alpha} \Phi^{-\alpha+1} \Pi_1 \Pi_2  + \frac{a_1}{\alpha^2} \Phi_1^{-\alpha} \Phi_2 (\Pi_2)^2 + \alpha \Phi_1^{\alpha-1} \vec{\nabla} \Phi_1 . \vec{\nabla} \Phi_2  \nonumber \\
&-& a_1 \Phi_1^{\alpha-2} \Phi_2 (\vec{\nabla} \Phi_1)^2 + a_2 m^\beta \Phi_1^\gamma \Phi_2 + C.C.
\end{eqnarray}
where $C.C.$ in the above equation means the complex conjugate of the previous terms in the equation. The values of $a_{1,2}$ and $\beta$ are given in this case by eqs.(\ref{beta}-\ref{relation2}).

\subsection{The Real Case (I): Lagrangian and Hamiltonian}
\label{real1lagangian}

In order to find a Lagrangian density to the real solution of case I given in subsection \ref{real1.1} we propose an ansatz to this Lagrangian. We compute the Euler-Lagrange equation and we compare with the solution we have found in subsection \ref{real1.1}. Let us consider the following Lagrangian
\begin{eqnarray}
\label{ansatzreallagrangian1}
\mathcal L = \kappa_1 \Phi(x)^{\alpha-1} \partial_\mu\Phi(x) \partial^\mu\Phi(x) + \kappa_2 m^2 \Phi(x)^{\alpha+1}
\end{eqnarray}
We remind that in this case the relativistic dispersion relation give $\alpha = \gamma$ the constant $c$ is arbitrary and $\beta = 2$. Moreover, as $b \neq 0$ eq.(\ref{eqreal1.1})  implies that $a_1=\alpha(1-\alpha \theta)/\theta$ and from eq.(\ref{eqreal1.2}) we get $a_2 = \alpha \theta b^2$. From the above eq.(\ref{ansatzreallagrangian1}) we have
\begin{eqnarray}
\label{elreal1}
\frac{\partial \mathcal L}{\partial \Phi} &=& \kappa_1 (\alpha-1) \Phi^{\alpha-2} \partial_\mu \Phi \partial^\mu \Phi + \kappa_2 (\alpha+1) m^2 \Phi^\alpha \\
\label{elreal2}
\partial_\mu\frac{\partial \mathcal L}{\partial \partial_\mu\Phi} &=& 2 \kappa_1 (\alpha-1) \Phi^{\alpha-2} \partial_\mu \Phi \partial^\mu \Phi + 2 \kappa_1 \Phi^{\alpha-1} \partial_\mu \partial^\mu\Phi
\end{eqnarray}
Using the above eqs.(\ref{elreal1} - \ref{elreal2}) the Euler-Lagrange equations for this case is
\begin{eqnarray}
\label{elreal1f}
2 \kappa_1 \Phi^{\alpha-1} \partial_\mu\partial^\mu \Phi + \kappa_1(\alpha-1) \Phi^{\alpha-2} \partial_\mu \Phi\partial^\mu \Phi - \kappa_2 (\alpha+1) m^2 \Phi^\alpha=0
\end{eqnarray}
Comparing with eq.(\ref{ansatz1}) we have the following equations $\textnormal{(i)} 2\kappa_1=\alpha$, $\textnormal{(ii)} \kappa_1 (\alpha-1) = a_1 + \alpha(\alpha-1)$ and $\textnormal{(iii)} -\kappa_2(\alpha+1)=a_2$. Eqs. (i) and (ii) together gives $a_1=-\alpha(\alpha-1)/2$ that together with $a_1=\alpha(1-\alpha \theta)/\theta$ gives us an additional condition to $\theta$, namely $\theta = 2/(1+\alpha)$. Using above equations (i), (iii) and this value of $\theta$ the Lagrangian density for this case becomes: 
\begin{eqnarray}
\label{reallagrangian1final}
\mathcal L =\frac{\alpha}{2} \Phi^{\alpha-1} \partial_\mu \Phi \partial^\mu \Phi - \frac{2\alpha b^2}{(\alpha+1)^2} m^2 \Phi^{\alpha+1}
\end{eqnarray}

For this case the canonically conjugate momentum to $\Phi$ is $\Pi = \partial\mathcal L/ \partial \dot\Phi = \alpha \Phi^{\alpha-1} \dot\Phi$ and the Hamiltonian is
\begin{eqnarray}
\label{hamiltonianreal1.1}
\mathcal H = \frac{\alpha}{2} \Phi^{\alpha-1} (\dot\Phi)^2 +  \frac{\alpha}{2} \Phi^{\alpha-1} \, (\vec{\nabla} \Phi)^2 + \frac{\alpha \theta b^2}{\alpha+1} m^2 \Phi^{\alpha+1}
\end{eqnarray}
But since $\Pi^2 = \alpha^2  \Phi^{2(\alpha-2)} \dot\Phi^2$  finally the Hamiltonian can be written as
\begin{equation}
\label{finalformhamiltonianreal1}
\mathcal H =  \frac{1}{2\alpha} \Phi^{-\alpha+1} \Pi^2 + \frac{\alpha}{2}  \Phi^{\alpha-1} \, (\vec{\nabla} \Phi)^2 + \frac{\alpha \theta b^2}{\alpha+1} m^2 \Phi^{\alpha+1} .
\end{equation}

\subsection{The Real Case (II): Lagrangian and Hamiltonian}
\label{real2lagangian}

In subsection \ref{real2.1} we have found $a_1=- \alpha^2$, $a_2 = c^{\alpha -\gamma} \alpha  \theta b^2$, $\alpha -\gamma= 2/\theta$ and $\beta = 2 + \delta (\alpha -\gamma)$. Thus, in this case the generalized equation, eq.(\ref{ansatz1}), is:
\begin{eqnarray}
\label{eqreal2}
\Phi^{\alpha-1} \partial_\mu \partial^{\mu} \Phi -\Phi^{\alpha-2} \partial_\mu \Phi \partial^\mu \Phi +
c^{2/\theta} \theta b^2 m^\beta \Phi^\gamma = 0
\end{eqnarray}
Notice that, this is the only case that we do not recover in any limit the standard Klein-Gordon equation.

Proceeding as in subsection \ref{real1lagangian}, i.e., choosing the simple Lagrangian to describe this system
\begin{eqnarray}
\label{ansatzreallagrangian2}
\mathcal L = \kappa_1  \Phi^{\alpha} \partial_\mu \partial^{\mu} \Phi + 
\kappa_2 \Phi^{\alpha-1} \partial_\mu \Phi \partial^\mu \Phi +
\kappa_3 m^\beta \Phi^{\gamma+1}
\end{eqnarray}
computing the Euler-Lagrange equation and comparing with eq.(\ref{ansatz1})  we find solution only for two space-time dimensions. Thus, in order to obtain a general solution we are going to introduce an auxiliary field as was used in the cases of the $q$-Schr\"odinger, the complex $q$-Klein-Gordon equations presented in \cite{q-3eq}
and in the complex case of subsection \ref{complexlagangian}.
Then, we introduce an auxiliary field $\Phi_2$ and the following Lagrangian:
\begin{eqnarray}
\label{ansatzreallagrangian3}
\mathcal L =  \Phi_1^{\alpha-1} \partial_\mu \Phi_2 \partial^{\mu} \Phi_1 + 
\alpha \Phi_2 \Phi_1^{\alpha-2} \partial_\mu \Phi_1 \partial^\mu \Phi_1 -
p \Phi_2 \Phi_1^{\gamma}
\end{eqnarray} 
where in the above equation $p \equiv c_1^{2/\theta} \theta b^2 m^\beta$. Notice that, the Euler-Lagrangian for $\Phi_2$ of Lagrangian eq.(\ref{ansatzreallagrangian3}) gives eq.(\ref{eqreal2}). We are going now to compute the Euler-Lagrange equation for $\Phi_1$ of Lagrangian eq.(\ref{ansatzreallagrangian3}). It is easy to see that
\begin{eqnarray}
\label{elreal2.1}
\frac{\delta \mathcal L}{\delta \Phi_1} = 
(\alpha-1)  \Phi_1^{\alpha-2} \partial_\mu \Phi_2  \partial^\mu \Phi_1 +
\alpha (\alpha-2) \Phi_2 \Phi_1^{\alpha-3} \partial_\mu \Phi_1 \partial^\mu \Phi_1 - \gamma p \Phi_2 \Phi_1^{\gamma-1} 
\\
\partial_\mu\frac{\delta \mathcal L}{\delta \partial_\mu \Phi_1} = (3 \alpha-1)  \Phi_1^{\alpha-2} \partial_\mu \Phi_2 \partial^\mu \Phi_1 +
2 \alpha \Phi_2 \Phi_1^{\alpha-2}  \partial^\mu \partial_\mu \Phi_1 +
\Phi_1^{\alpha-1} \partial_\mu \partial^\mu \Phi_2  \nonumber \\
+ 2 \alpha (\alpha-2) \phi_1^{\alpha-3} \Phi_2  \partial_\mu \Phi_1 \partial^\mu \Phi_1
\end{eqnarray}
Thus the Euler-Lagrange equation for the field $\Phi_1$ is
\begin{eqnarray}
\label{eqauxreal2}
\Phi_1^{\alpha-1} \partial_\mu \partial^\mu \Phi_2 + 
2 \alpha \Phi_1^{\alpha-2} \partial_\mu \Phi_1 \partial^\mu \Phi_2 + 
\left[ \alpha (\alpha-2) \phi_1^{\alpha-3} \partial_\mu \Phi_1 \partial^\mu \Phi_1  \right. \nonumber  \\ 
\left. + 2 \alpha \Phi_1^{\alpha-2}  \partial_\mu \partial^\mu \Phi_1 + 
\gamma p \Phi_1^{\gamma-1} \right] \Phi_2 =0
\end{eqnarray}
As in the complex case, eq.(\ref{auxeq1}), this equation is linear in $\Phi_2$ and in order to solve it we must introduce the value of $\Phi_1$ which is solution of eq.(\ref{eqreal2}). Choosing $\Phi_1(z) = c_1 m^\delta \cos(b z)^\theta$ and its derivatives into eq.(\ref{eqauxreal2}) and writing $\Phi_2 = c_2 m^\delta g(z)$ the equation becomes
\begin{eqnarray}
\cos(b z)^{2 \theta } g^{\prime \prime}(z) - 
2 b \alpha  \theta  \cos(b z)^{-1+2 \theta } \sin(b z) g^{\prime}(z) \nonumber \\
-\frac{1}{2} b^2 \cos(b z)^{-2+2 \theta } \left(4+2 \alpha  \theta -\alpha ^2 \theta ^2+\alpha ^2 \theta ^2 \cos(2 b z)\right) g(z) =0
\end{eqnarray}
It can be verified that the solution to this equation is:
\begin{eqnarray}
\label{solutioncase2}
g(z)=\chi_1 \cos(b z)^{-1-\alpha  \theta } \sin(b z) + \chi_2 \cos(b z)^{-1-\alpha  \theta }  \nonumber \\
\left( \cos(b z)-2 \sin(b z) \tan^{-1} \left[ \cos(bz)/(-1+\sin(b z) \right]  \right)
\end{eqnarray}
where $\chi_{1,2}$ are two arbitrary constants. Thus, the solution to eq.(\ref{eqauxreal2}) is $\Phi_2 = c_2 m^\delta g(z)$ where $g(z)$ is given by eq.(\ref{solutioncase2}),  $\delta= (\nu-2)/(1+\alpha)$ and $\gamma = \alpha-2/\theta$.

The canonically conjugate momenta to $\Phi_{1,2}$ are in this case $\Pi_1 = \frac{\partial \mathcal L}{\partial \dot\Phi_1} =  \Phi_1^{\alpha-1} \dot\Phi_2 + 2 \alpha \Phi_1^{\alpha-2} \Phi_2 \dot\Phi_1$ and $\Pi_2 = \frac{\partial \mathcal L}{\partial \dot\Phi_2} =  \Phi_1^{\alpha-1} \dot\Phi_1 $. Thus, $\mathcal H = \Pi_1 \dot\Phi_1 + \Pi_2 \dot\Phi_2 - \mathcal L$ and we get for the Hamiltonian:
\begin{eqnarray}
\label{hamiltonianreal2}
\mathcal H = \Phi_1^{\alpha-1} \dot\Phi_1 \dot\Phi_2  + \alpha \Phi_1^{\alpha-2} (\dot\Phi_1)^2 \Phi_2 +  \Phi_1^{\alpha-1} \vec{\nabla} \Phi_1 . \vec{\nabla} \Phi_2 +
\alpha \Phi_1^{\alpha-2} (\vec{\nabla} \Phi_1)^2  \Phi_2 +  \nonumber  \\  
p\Phi_2 (\Phi_1)^{\gamma}
\end{eqnarray}
which can be rewritten as
\begin{eqnarray}
\label{hamiltonianreal3}
\mathcal H =   \Phi_1^{-\alpha+1} \Pi_1 \Pi_2 -  \alpha  \Phi_2 \Phi_1^{-\alpha}  \Pi_2^2 + 
\Phi_1^{\alpha-1} \vec{\nabla} \Phi_1 . \vec{\nabla} \Phi_2 +
\alpha \Phi_1^{\alpha-2} (\vec{\nabla} \Phi_1)^2  \Phi_2  +  \nonumber  \\ 
p \Phi_2 (\Phi_1)^{\gamma}
\end{eqnarray}
where we recall that $p \equiv c_1^{2/\theta} \theta b^2 m^\beta$ and  $\gamma = \alpha-2/\theta$.

\section{Lorentzian Soliton}
\label{lorentzian}

We are going to show in this section that a subclass of the class of complex solutions to the nonlinear generalized Klein-Gordon equation has a spatial one-dimensional Lorentzian solitonic solution. For this purpose we consider the Hamiltonian in eq.(\ref{hamiltonian1}) with the solutions $\Phi_1(z) = m^\delta c \, e_q(z)$ and $\Phi_2$ given in eq.(\ref{solutionphi2}). Moreover recall that $z \equiv i k.x = i (\omega t - \vec k. \vec x)$ and the solution we shall find has finite energy only for one-spatial dimension. For simplicity, from now on we omit the dependence on space-time when we write the Hamiltonian density, for instance, the $q$-exponential $e_q(z)$ will be written simply as $e_q$. Thus, computing derivatives of the solutions $\Phi_{1,2}$ we arrive at:
\begin{eqnarray}
\label{density1}
\frac{\mathcal H}{c_1^\alpha c_2 m^{\delta(\alpha+1)}} = \left[ (a_1-\alpha r_1)(\omega^2+\vec{k}^2)+ \left( \alpha q+a_1+ \alpha(\alpha-1)\right) m^2 \right] \kappa_1 e_q^{\Delta_1} + \nonumber \\
\,\,\,\,\,\,\,\,\,\,\,\,+ \left[ (a_1-\alpha r_2)(\omega^2+\vec{k}^2)+ \left( \alpha q+a_1+ \alpha(\alpha-1)\right) m^2 \right] \kappa_2 e_q^{\Delta_2} + C.C.
\end{eqnarray}
where $\Delta_{1,2} \equiv r_{1,2}+2q+\alpha-2$, giving $\Delta_1 = 0$, 
\begin{eqnarray}
\label{deltavalue}
\Delta_2 = 3(q-1)+2(a_1/ \alpha)+ 2\alpha
\end{eqnarray}
and $r_2-r_1 = \Delta_2$. Where eqs.(\ref{rvalue1}-\ref{rvalue2})  were used.

We introduce to the Hamiltonian density an interaction term of the type $\lambda \Phi_2 \Phi_1^\gamma$ which leads us to
\begin{equation}
\frac{\mathcal H + \lambda \left(\kappa_1 e_q^{\Delta_1} + \kappa_2 e_q^{\Delta_2} \right)}{c_1^\alpha c_2 m^{\delta(\alpha+1)}} 
\end{equation}
If we choose for $\lambda$ the value
\begin{equation}
\label{lambda}
\lambda = - \left[ (a_1-\alpha r_1)(\omega^2+\vec{k}^2)+ \left( \alpha q+a_1+ \alpha(\alpha-1)\right) m^2 \right] 
\end{equation}
we find
\begin{equation}
\label{densitysolution}
\frac{\mathcal H + \lambda \left(\kappa_1 e_q^{\Delta_1} + \kappa_2 e_q^{\Delta_2} \right)}{c_1^\alpha c_2 m^{\delta(\alpha+1)}} = - \alpha \Delta_2(\omega^2+k^2)  \kappa_2 e_q^{\Delta_2} +C.C.
\end{equation}
The choice of this value of $\lambda$ annihilates the $\kappa_1$ sector in eq.(\ref{density1}). As $\Delta_1=0$ this part of the density would produce a divergent result of the integration even for one spatial dimension, thus the introduction of this particular interaction term can be interpreted as a classical renormalization of energy. Now, choosing for $a_1$ the value
\begin{equation}
\label{alphacomplex}
a_1 = \alpha (1-q) -\alpha^2
\end{equation}
we have $\Delta_2 = q - 1$ and in this case the real part of $e_q^{\Delta_2}$ is $\Re \left[ e_q^{q-1}\right] = 1/(1+(1-q)^2 (kx-\omega t)^2)$. Finally, 
\begin{equation}
\label{lorentziansoliton}
\frac{\mathcal H + \lambda \left(\kappa_1 e_q^{\Delta_1} + \kappa_2 e_q^{\Delta_2} \right)}{c_1^\alpha c_2 m^{\delta(\alpha+1)}} =  \frac{ 2 \alpha (\omega^2+k^2)\kappa_2(1-q)}{1+(1-q)^2 (kx-\omega t)^2}
\end{equation}
Since the integral of the above term is $2 \alpha(\omega^2+k^2)\kappa_2(1-q) \pi/|1-q|$ the energy is finite.  By a simple computation we can show that the height of the peak of the density in eq.(\ref{lorentziansoliton}) is independent of time with value $2 \alpha(\omega^2+k^2)\kappa_2(1-q)$, the shape of the density does not change with time and then we call this solution a Lorentzian soliton. It is important to mention that there is no limit of this Lorentzian soliton solution to the standand Klein-Gordon equation since it appears in a different sector. As commented in paragraph below eq.(\ref{rvalue2}) the limit to the standard Klein-Gordon solution is obtained by taking $\kappa_2=0$ while the solitonic solution is obtained for $\kappa_2 \neq 0$.

Choosing for $\kappa_2$ 
\begin{eqnarray}
\label{kappa2}
|\kappa_2| &=& \frac{1}{\omega^2+k^2} \nonumber \\
\sgn(\kappa_2) &=& \sgn(\alpha) \sgn(1-q)
\end{eqnarray}
where $\sgn(x)$ is the signum function of $x$, we have for the Lorentzian soliton
\begin{eqnarray}
\label{solitoncomplexconstantsfinal}
\frac{\mathcal H + \lambda \left(\kappa_1 e_q^{\Delta_1} + \kappa_2 e_q^{\Delta_2} \right)}{c_1^\alpha c_2 m^{\delta(\alpha+1)}} = 
\frac{2 |q -1| |\alpha| }{1  + (1-q)^2  \hat z^2}
\end{eqnarray}
where $\hat z \equiv  \omega t - k x$ and the integral of the expression in eq.(\ref{solitoncomplexconstantsfinal}) gives $2 |\alpha| \pi$. 
Notice that in order to recover the $q$-Klein-Gordon model in \cite{q-3eq} we have to choose $a_1=0$ and $\alpha = 1$. In this case,  eq. (\ref{alphacomplex}) implies a soliton solution for $q=0$. 

In order to eliminate the sector proportional to $\kappa_1$ in eq.(\ref{density1}), which is responsible for a divergent energy, one could be tempted to choose $\kappa_1 =0$. However, in order to obtain a solitonic solution in the second sector $\kappa_2 \neq 0$ it is necessary to have $\Delta_2 = q-1$ which gives eq.(\ref{alphacomplex}). But this value of $a_1$ in eq.(\ref{alphacomplex}) cancels also the term proportional to $\kappa_2$ thus preventing the existence of the soliton solution. Then, in order to have the Lorentzian soliton it is necessary to perform the above classical renormalization of the energy.

\begin{figure}
\begin{center}
\includegraphics[width=4in]{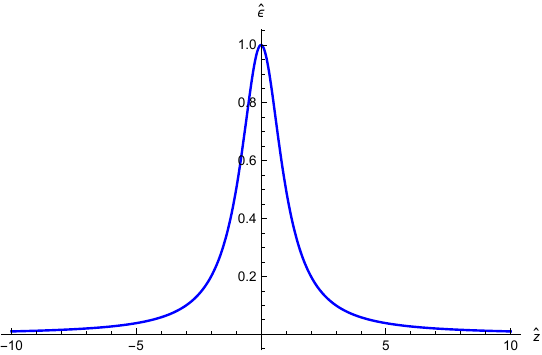}
\caption{ Solitary wave behaviour of the density of energy. Above is the plot of $\hat{\epsilon}$ in eq.(\ref{solitoncomplexconstantsfinal}), where $\hat{\mathcal{E}} \equiv \frac{\mathcal H + \lambda \left(\kappa_1 e_q^{\Delta_1} + \kappa_2 e_q^{\Delta_2} \right)}{c_1^\alpha c_2 m^{\delta(\alpha+1)}2 |q -1| | \alpha |}$ for the value $q=2$. Note that, $\hat z$ is the horizontal and $\hat{\mathcal{E}}$ is the vertical axis.}
\label{density}
\end{center}
\end{figure}

In Figure \ref{density}  we show the solitonic behavior of $\hat{\mathcal{E}} \equiv \frac{\mathcal H + \lambda \left(\kappa_1 e_q^{\Delta_1} + \kappa_2 e_q^{\Delta_2} \right)}{c_1^\alpha c_2 m^{\delta(\alpha+1)} 2 |q -1| |\alpha|} $ for the value $q=2$.  

\vspace{2cm}

\section{Final Comments}
\label{final}

We have derived in this paper in sections \ref{generalization} - \ref{lagangian/hamiltonian} three classes of nonlinear generalized Klein-Gordon models with fields having mass dimension depending on a real parameter $\alpha$ as $\delta=2/(1+\alpha)$ where $\delta$ is the mass dimension of the field.  models in these three classes have travelling-wave solutions and satisfy the relativistic dispersion relation. 
 In subsection \ref{complex} we derived the class of nonlinear generalized Klein-Gordon equations having complex solutions. In subsection \ref{solutionreal} we have shown the there are two classes of solutions for real fields. In one class of real fields there is a limit where we recover the standard Klein-Gordon equation, instead the other class there is no such limit. In section \ref{lagangian/hamiltonian} the compute the Lagrangian and Hamiltonian for the three classes.

The fields in the models of these three classes could in principle have he mass dimension varying from zero to one and this allow us to construct interaction terms different from $\lambda \Phi^4$ with coupling constants having mass dimensions positive or zero. Thus, from what we know about perturbation theory these interacting models are candidates to be renormalized or super-renormalized ones. Of course, in order to check this possibility one should firstly quantize these models, develop perturbation theory for them and prove their renormalizability or super renormalizability. Moreover, other issues, as for instance unitarity and locality among others, must be checked as, for instance, higher derivatives field theory teach us. Also, symmetries as for instance parity must also be checked in the quantized model.

In section \ref{lorentzian} we have shown that a subclass of the class of complex solutions to the nonlinear generalized Klein-Gordon equation has a spatial one-dimensional Lorentzian solitonic solution. 
It is also worth mentioning that it seems to us likely he model we call Case (I) in subsection \ref{real1.1} can have a solution generalizing the sine-gordon soliton and therefore we believe it is worth investigating.

\vspace{2cm} 
\noindent 
{\bf Acknowledgments:}   The authors thank Constantino Tsallis, Francesco Toppan, Fernando D. Nobre and Sebasti\~ ao Alves Dias  for discussions.

\end{document}